\documentclass[RNAAS]{aastex63}
\graphicspath{{./}{figures/}}
\usepackage{rotating}
\usepackage{lineno}
\begin{document}

\def\teff{$\rm T_{\rm eff}$}
\def\kms{$\rm km s^{-1}$}
\def\msun{$M_\odot$}

\title{A new set of chisels for Galactic archaeology: Sc, V and Zn as taggers of accreted globular clusters
\footnote{Based on observations collected at the ESO-VLT under programs 073.D-0211, 079.B-0721, 193.D-0232}}

\author{A. Minelli}
\affiliation{Dipartimento di Fisica e Astronomia {\sl Augusto Righi}, Universit\`a degli Studi di Bologna, Via Gobetti 93/2, I-40129 Bologna, Italy}
\affiliation{INAF - Osservatorio di Astrofisica e Scienza dello Spazio di Bologna, Via Gobetti 93/3, I-40129 Bologna, Italy}

\author{A. Mucciarelli}
\affiliation{Dipartimento di Fisica e Astronomia {\sl Augusto Righi}, Universit\`a degli Studi di Bologna, Via Gobetti 93/2, I-40129 Bologna, Italy}
\affiliation{INAF - Osservatorio di Astrofisica e Scienza dello Spazio di Bologna, Via Gobetti 93/3, I-40129 Bologna, Italy}

\author{D. Massari}
\affiliation{INAF - Osservatorio di Astrofisica e Scienza dello Spazio di Bologna, Via Gobetti 93/3, I-40129 Bologna, Italy}
\affiliation{University of Groningen, Kapteyn Astronomical Institute, NL-9747 AD Groningen, The Netherlands}

\author{M. Bellazzini}
\affiliation{INAF - Osservatorio di Astrofisica e Scienza dello Spazio di Bologna, Via Gobetti 93/3, I-40129 Bologna, Italy}

\author{D. Romano}
\affiliation{INAF - Osservatorio di Astrofisica e Scienza dello Spazio di Bologna, Via Gobetti 93/3, I-40129 Bologna, Italy}

\author{F. R. Ferraro}
\affiliation{Dipartimento di Fisica e Astronomia {\sl Augusto Righi}, Universit\`a degli Studi di Bologna, Via Gobetti 93/2, I-40129 Bologna, Italy}
\affiliation{INAF - Osservatorio di Astrofisica e Scienza dello Spazio di Bologna, Via Gobetti 93/3, I-40129 Bologna, Italy}

\begin{abstract}
Chemical tagging is a powerful tool to reveal the origin of stars and globular clusters (GCs), especially when dynamics alone cannot provide robust answers. So far, mostly $\alpha$- and neutron capture elements have been used to distinguish stars born in the Milky Way (MW) from those born in external environments such as that of dwarf galaxies.
Here, instead, we use iron-peak elements abundances to investigate the origin of a sample of metal-rich globular clusters. By homogeneously analyzing high-resolution UVES spectra of giant stars belonging to four metal-rich GCs (namely NGC 5927, NGC 6388, NGC 6441, NGC 6496), we find that while the $\alpha$-elements Si and Ca have similar abundance ratios for all the four GCs, and Ti and neutron capture elements (La, Ba and Eu) only show a marginal discrepancy, a stark difference is found when considering the abundances of some iron-peak elements (Sc, V and Zn). In particular, NGC 6388 and NGC 6441 have abundance ratios for these iron-peak elements significantly lower (by $\sim$ 0.5 dex) than those measured in NGC 5927 and NGC 6496, which are clearly identified as born in-situ MW clusters through an analysis of their orbital properties. These measurements indicate that the environment in which these clusters formed is different, and provide robust evidence supporting an accreted origin from the same progenitor for NGC 6388 and NGC 6441.
\end{abstract}

\section{Introduction}

According to the generally accepted $\Lambda$-CDM cosmological model, large galaxies that we observe today were formed from the merging of small structures \citep{White1978, Helmi2018}. The Milky Way (MW) is an excellent example of this assembly mechanism, since in the past it has experienced several merger events, many of which have been recently discovered thanks to the advent of the Gaia mission (e.g., Gaia-Enceladus-Sausage, Sequoia, Thamnos, see \cite{Helmi2020} for a comprehensive review).
During this assembly process, the MW accreted both field stars and globular clusters (GCs). In particular, about 50\% - 60\% of its current population of GCs has likely been accreted from different external progenitors, while the rest has likely formed in-situ \citep{Massari2019, Forbes2020}.
So far, the accreted or in-situ origin of the GCs has been primarily assessed by using their dynamics, coupled with information on their age-metallicity relation \citep{Kruijssen2019, Massari2019}. However, the dynamical properties of some GCs do not allow a clear-cut classification.

Chemical tagging \citep{Freeman2002} is 
a powerful tool to reveal the origin of stars by means of their chemical patterns. In particular, it has been shown both theoretically \citep[e.g.][]{Matteucci1990}  and observationally \citep[e.g.][]{Helmi2018, Fernandez2018}, that abundance of $\alpha$-elements is an efficient tool to distinguish stars born in the MW from those born in dwarf galaxies. Furthermore, the slow neutron capture elements were observed to be enhanced in dwarf galaxies with respect to MW stars of similar metallicity \citep[see e.g.][]{Tolstoy2009}.
\\
\cite{Minelli2021} proposed to use the chemical abundance ratios of some iron-peak elements, namely Sc, V and Zn, 
as diagnostics to identify possible extra-galactic stars in the metal-rich regime ([Fe/H] $>$ --1 dex). 
In a thorough and homogeneous comparison of chemical compositions between the Large Magellanic Cloud (LMC), the Sagittarius (Sgr) dwarf spheroidal and the MW, they found that the largest difference between LMC/Sgr and MW stars occurs in [V/Fe] and [Zn/Fe], reaching up to 0.5/0.7 dex for the latter ratio.

These usually poorly explored abundance ratios are, thus, able to distinguish stars formed in low star-formation rate environments, like those of dwarf galaxies, in the metal-rich regime, where more commonly investigated abundance ratios (like the explosive $\alpha$-elements or neutron capture elements) lose their sensitivity as a proxy of different stellar birth places.

In this Letter we apply the tools proposed in \cite{Minelli2021} for chemical tagging to four metal-rich GCs, namely NGC~5927, NGC~6496, NGC~6388 and NGC~6441. These GCs have  similar metallicities ([Fe/H] $\sim$ -- 0.5 dex) and they are thus located in the metallicity range 
where the iron-peak element abundance ratios should exhibit the largest discrepancy in case they have a different origin \citep{Minelli2021}. According to their dynamical properties, the first two have been clearly identified as in-situ clusters \citep{Massari2019}. On the other hand, the other two seem to share an accreted origin, but their orbital properties make their classification more uncertain \citep[see][]{Massari2019, Kruijssen2020}.
These two clusters are usually associated each other, in particular 
because they exhibit extended blue horizontal branches \citep{Rich1997}, 
despite of their high metallicity, suggesting an high He content \citep{Bellini2013}.

\section{SPECTROSCOPIC DATASETS}
All the spectra have been acquired with the multi-object spectrograph UVES-FLAMES \citep{Pasquini2002} mounted at the Very Large Telescope of ESO, using the grating 580 Red Arm CD\#3, which provides a spectral resolution of R=~47000 and a spectral coverage between 4800 and 6800 \AA. They have been reduced with the dedicated ESO pipelines\footnote{http://www.eso.org/sci/software/pipelines/}, including bias subtraction, flat-fielding, wavelength calibration, spectral extraction and  order  merging. For each individual spectrum, the sky background has been subtracted, 
using the spectra obtained observing empty sky regions.
\\
Considering the high luminosity/low temperature and high metallicity of the observed stars, 
we check for the presence of TiO molecular bands that can affect the derived chemical abundances and we exclude the contaminated spectra.\\
The targets of our analysis are four GCs. Their data were collected as follows:
\begin{itemize}
\item \textit{NGC 5927} -- NGC 5927 is a disky MW GC \citep[according to the classification adopted by][disk clusters have the maximum height from the disk $Z_{max}<5$ kpc and the orbital circularity $circ<0.5$]{Massari2019}.
It has a metallicity of [Fe/H] = -- 0.47 $\pm$ 0.02 dex \citep{Mura2018} and a mass of $2.75 \pm 0.02 \times 10^5$ \msun \citep[the value is taken from the current latest version of the  globular cluster database by Holger Baumgardt, see][]{Baumgardt2018}.
The dataset for this GC is composed of five red giant branch (RGB) stars, observed under the ESO-VLT program 079.B-0721 (PI: Feltzing).

\item \textit{NGC 6441} -- 
This cluster has a metallicity of [Fe/H] = -- 0.39 dex $\pm$ 0.04 \citep{Gratton2006}, a mass of $1.32 \pm 0.01 \times 10^6$ \msun\  \citep{Baumgardt2018} and despite its orbit currently place it in the Galactic Bulge, it likely has an accreted origin according to \cite{Massari2019}. Among the four members identified by \cite{Gratton2006}, we include in our analysis only the two giant stars observed under the ESO-VLT program: 073.D-0211 (PI: Carretta), whose spectra are not contaminated by TiO molecular bands.

\item \textit{NGC 6388} -- This cluster has a similar orbit compared to that of NGC 6441, yet \cite{Massari2019} classify it as an in-situ Bulge GCs (these authors defines as bulge clusters those placed on highly bound orbits, with apocenter $apo<3.5$ kpc). It has a mean metallicity of [Fe/H] = -- 0.44 $\pm$ 0.01 dex \citep{Carretta2007} and a mass of $1.25 \pm 0.01 \times 10^6$ \msun \citep{Baumgardt2018}. Among NGC 6388 stars observed under the ESO-VLT program: 073.D-0211 (PI: Carretta), we analyzed the four giants that are cluster members according to their radial velocity \citep{Carretta2007} and whose spectra were not contaminated by TiO molecular bands.

\item \textit{NGC 6496} -- Just like NGC 5927, NGC 6496 is a disky MW GC. It has a metallicity of [Fe/H] = -- 0.46 $\pm$ 0.07 dex derived from low resolution spectra \citep{Carretta2009} and a mass of $6.89 \pm 0.73 \times 10^4$ \msun \citep{Baumgardt2018}. This dataset includes five RGB stars belonging to this GC observed in the contest of the ESO-MIKiS survey \citep{Ferraro2018}, Large Programme 193.D-0232 (PI: Ferraro). The member stars are selected according to their radial velocity.  
\end{itemize}

The elements we focus on in our investigation are generally not affected by the chemical pecularities associated to the so-called phenomenon of multi-populations in GCs \citep{Bastian2018}. The only possible exception is Sc, which shows possible variations in massive GCs \citep{Carretta2021}. NGC6441 and NGC6388 are indeed massive, but they do not show Sc variations according to the quoted analysis.
 
\section{Analysis}

The four target GCs are characterized by large values of color excess and differential reddening that make the atmospheric parameters of individual stars uncertain when derived from the photometry. 
Thanks to the large number of Fe~I lines available in the UVES spectra, 
effective temperatures (\teff) can be easily derived by imposing the excitation equilibrium. As discussed by \cite{Mucciarelli20}, for metal-rich giant stars spectroscopic temperatures are consistent with the photometric temperatures and the method  can be adopted safely \citep[at variance with the metal-poor stars where the spectroscopic temperatures are biased and systematically under-estimated, as shown in Fig. 9 of][]{Mucciarelli20}.

The surface gravity (logg) is derived by using theoretical isochrones computed with the Dartmouth Stellar Evolution Database \citep{isocrone}, adopting for each GC an isochrone with appropriate age \citep{Forbes2010} and chemical mixture
(we started with the literature value of [Fe/H] and [$\alpha$/Fe], adapting in each interaction their value to the results of our analysis).

Finally, the microturbolent velocities ($\xi$) of the stars are derived spectroscopically, by minimizing the slope between the abundances from Fe I lines and the reduced equivalent widths.

Abundances of Si, Ca, Ti and Fe have been derived from the measured equivalent widths (EWs) of unblended lines using the code GALA \citep{Mucciarelli2013}. The EWs have been measured with DAOSPEC \citep{StetsonPancino2008} through the wrapper 4DAO \citep{4dao}. A line-by-line inspection has been performed in order to check the continuum location and the best-fit for each individual line.

The chemical abundances for the species for which only blended lines (Sc, V, Ba, La and Eu) or transitions located in noisy/complex spectral regions (Zn) are available, have been derived with our own
code SALVADOR that performs a $\chi^2$-minimization between the observed line and a grid of suitable synthetic spectra
calculated on the fly using the code SYNTHE \citep{Kurucz2005}.

We exclude from our analysis those elements (O, Na, Mg and Al) involved in the multiple population phenomenon \citep{Bastian2018}.

The procedure to select the lines used to derive the chemical abundances of the involved elements is described in \cite{Minelli2021}.
Atomic data for the selected lines are from the Kurucz/Castelli database, with more recent or more accurate data for some specific transition \citep[see][for additional references related to Fe, Si, Ca, Ti, Ba and Eu lines]{Mucciarelli2017}. 
Atomic data for Sc and V lines are from MFW e NBS \citep{Wiese1975, Martin1988}. For the Zn line at 4810 \AA\ we adopt the oscillator strength 
by \cite{Roederer2012}. Data for the La line at 6390 \AA\ are from \cite{Lawler2001}
Solar reference abundances are from \cite{Grevesse1998}, for consistency with \citet{Minelli2021}.

Errors in each abundance ratios have been calculated following the procedure described by \cite{Minelli2021} and propagating the uncertainties in astrophysical parameters into the chemical abundances.

\section{Results and discussion}
The objective of our analysis is to investigate whether the four GCs, all with a similar [Fe/H] $\sim$ -- 0.5 dex, show any differences in their elemental abundances, with particular focus on the iron-peak elements that have proven to be effective in distinguishing accreted from in-situ stars in this metal-rich regime. To do so, we homogeneously analyse high resolution spectra of RGB stars belonging to these Galactic GCs.
The abundances measured in individual stars, together with their atmospheric parameters, are listed in Table \ref{stelle}, instead the mean abundance ratios of the GCs for the analysed species are reported in Table \ref{medie}. 

\begin{sidewaystable*}\tiny
\centering
\caption{Atmospheric parameters, chemical abundance ratios and corresponding uncertainty for the individual target stars.}
\label{stelle}
\begin{tabular}{c|c|c|c|c|c|c|c|c|c|c|c|c|c|c|c|c|c|c|c|c|c|c|c}
star&T&log g&$\xi$&$[Fe/H]$&err&$[Si/Fe]$&err&$[Ca/Fe]$&err&$[Ti/Fe]$&err&$[Sc/Fe]$&err&$[V/Fe]$&err&$[Zn/Fe]$&err&$[Ba/Fe]$&err&$[La/Fe]$&err&$[Eu/Fe]$&err\\
\hline
\multicolumn{24}{c}{NGC 5927}\\
\hline
 5039161&4400&1.97&1.40&-0.42&0.04&+0.19&0.06&+0.10&0.08&+0.23&0.06&+0.28&0.05&+0.16&0.11&+0.07&0.07&+0.05&0.05&+0.08&0.06&+0.41&0.04\\
 5039423&4550&2.25&1.60&-0.47&0.05&+0.19&0.06&+0.06&0.07&+0.22&0.06&+0.32&0.05&+0.19&0.09&+0.22&0.09&+0.02&0.07&+0.22&0.04&+0.48&0.04\\
 5040219&4550&2.25&1.30&-0.51&0.04&+0.16&0.04&+0.10&0.06&+0.29&0.04&+0.36&0.04&+0.24&0.07&+0.32&0.07&+0.17&0.05&+0.28&0.04&+0.56&0.04\\
 5040282&4500&2.16&1.20&-0.44&0.04&+0.13&0.05&+0.20&0.06&+0.24&0.05&+0.33&0.04&+0.21&0.08&+0.40&0.07&+0.04&0.04&+0.24&0.04&+0.46&0.04\\
 5041223&4500&2.16&1.40&-0.47&0.05&+0.17&0.05&+0.07&0.07&+0.25&0.05&+0.38&0.05&+0.23&0.07&+0.30&0.07&+0.11&0.07&+0.24&0.04&+0.38&0.04\\
\hline
\multicolumn{24}{c}{NGC 6388}\\
\hline
   77599&4100&1.33&1.60&-0.49&0.09&+0.39&0.09&+0.07&0.12&+0.22&0.13&+0.01&0.05&-0.19&0.15&-0.07&0.09&-0.03&0.09&+0.16&0.05&+0.43&0.04\\
   83168&4150&1.42&1.50&-0.46&0.07&+0.19&0.09&-0.00&0.13&+0.15&0.13&+0.05&0.06&-0.17&0.17&-0.21&0.08&+0.08&0.08&+0.12&0.10&+0.31&0.04\\
  108895&4000&1.16&1.50&-0.51&0.03&+0.19&0.05&-0.04&0.09&+0.06&0.07&+0.02&0.04&-0.33&0.11&-0.12&0.06&+0.10&0.05&+0.07&0.04&+0.33&0.03\\
  110677&4000&1.16&1.50&-0.50&0.03&+0.19&0.06&-0.10&0.11&+0.03&0.10&-0.03&0.07&-0.32&0.13&-0.07&0.10&+0.05&0.06&+0.12&0.07&+0.27&0.04\\
\hline
\multicolumn{24}{c}{NGC 6441}\\
\hline
 7004463&3950&1.07&1.40&-0.48&0.03&+0.30&0.06&+0.16&0.11&+0.18&0.07&+0.11&0.10&-0.29&0.10&-0.38&0.14&+0.02&0.05&-0.03&0.05&+0.38&0.05\\
 7004487&4050&1.24&1.20&-0.59&0.04&+0.23&0.09&+0.11&0.09&+0.18&0.10&-0.05&0.09&-0.34&0.12&-0.60&0.14&-0.13&0.07&+0.04&0.07&+0.30&0.06\\
\hline
\multicolumn{24}{c}{NGC 6496}\\
\hline
 14&4150&1.48&1.30&-0.61&0.03&+0.31&0.04&+0.27&0.05&+0.34&0.05&+0.38&0.04&+0.25&0.12&+0.48&0.17&+0.27&0.06&+0.29&0.06&+0.52&0.05\\   
 17&4150&1.48&1.30&-0.64&0.03&+0.29&0.05&+0.20&0.07&+0.31&0.06&+0.41&0.05&+0.19&0.15&+0.21&0.17&+0.29&0.06&+0.32&0.06&+0.48&0.05\\    
 18&4150&1.48&1.40&-0.68&0.03&+0.27&0.05&+0.24&0.06&+0.29&0.05&+0.36&0.10&+0.17&0.14&+0.40&0.18&+0.08&0.11&+0.28&0.06&+0.48&0.06\\    
 26&4400&1.94&1.40&-0.60&0.04&+0.30&0.04&+0.05&0.04&+0.19&0.06&+0.36&0.05&+0.10&0.07&+0.12&0.09&+0.21&0.05&+0.20&0.05&+0.55&0.04\\    
159&4100&1.39&1.20&-0.65&0.03&+0.28&0.04&+0.23&0.05&+0.29&0.05&+0.37&0.04&+0.16&0.12&+0.23&0.09&+0.26&0.05&+0.21&0.04&+0.50&0.03\\    
\hline
\end{tabular}
\end{sidewaystable*}

\begin{table*}[!h]
\centering
\caption{Mean abundance ratios (and corresponding standard deviation) for the four target clusters.}
\label{medie}
\begin{tabular}{c|c|c|c|c|c|c|c|c}
&\multicolumn{2}{c|}{NGC 5927}&\multicolumn{2}{c|}{NGC 6388}&\multicolumn{2}{c|}{NGC 6441}&\multicolumn{2}{c}{NGC 6496}\\
element&mean&$\sigma$&mean&$\sigma$&mean&$\sigma$&mean&$\sigma$\\
\hline
$[Fe/H]$& -0.46&0.03&-0.49&0.02&-0.54&0.08&-0.64&0.03\\
$[Si/Fe]$&+0.17&0.02&+0.24&0.10&+0.27&0.05&+0.29&0.02\\
$[Ca/Fe]$&+0.11&0.06&+0.11&0.04&+0.14&0.04&+0.20&0.09\\
$[Ti/Fe]$&+0.25&0.03&+0.12&0.09&+0.18&0.00&+0.28&0.06\\
$[Sc/Fe]$&+0.33&0.04&+0.01&0.04&+0.03&0.11&+0.38&0.02\\
$[V/Fe]$& +0.21&0.03&-0.25&0.09&-0.32&0.04&+0.17&0.05\\
$[Zn/Fe]$&+0.26&0.12&-0.12&0.07&-0.49&0.16&+0.29&0.15\\
$[Ba/Fe]$&+0.07&0.06&+0.05&0.05&-0.05&0.10&+0.22&0.09\\
$[La/Fe]$&+0.21&0.08&+0.12&0.04&+0.01&0.05&+0.26&0.05\\
$[Eu/Fe]$&+0.46&0.07&+0.34&0.07&+0.34&0.06&+0.51&0.03\\
\hline	  	     	
\end{tabular}		
\end{table*}

\begin{figure}
\centering
\includegraphics[scale=0.29]{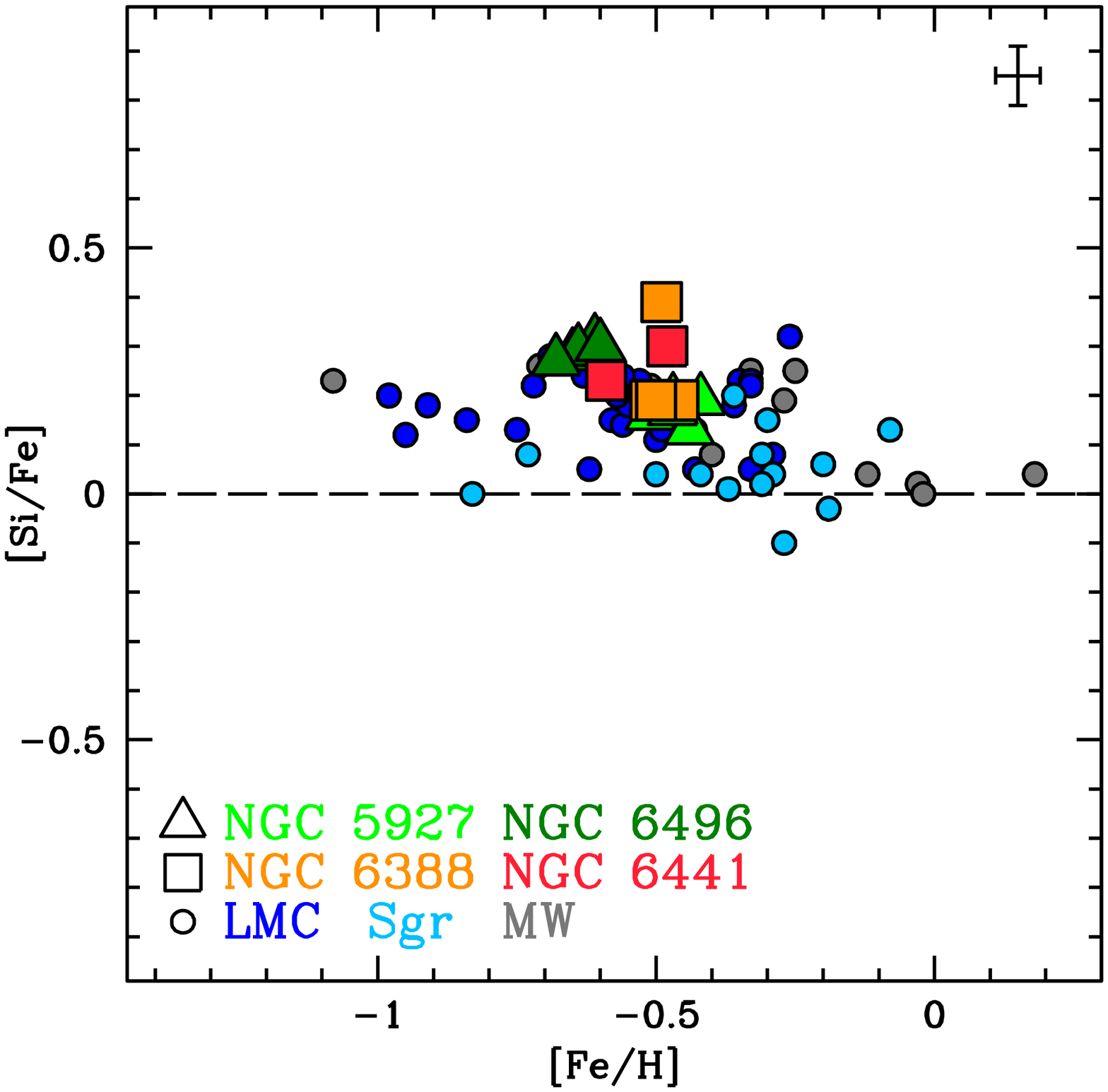}
\includegraphics[scale=0.29]{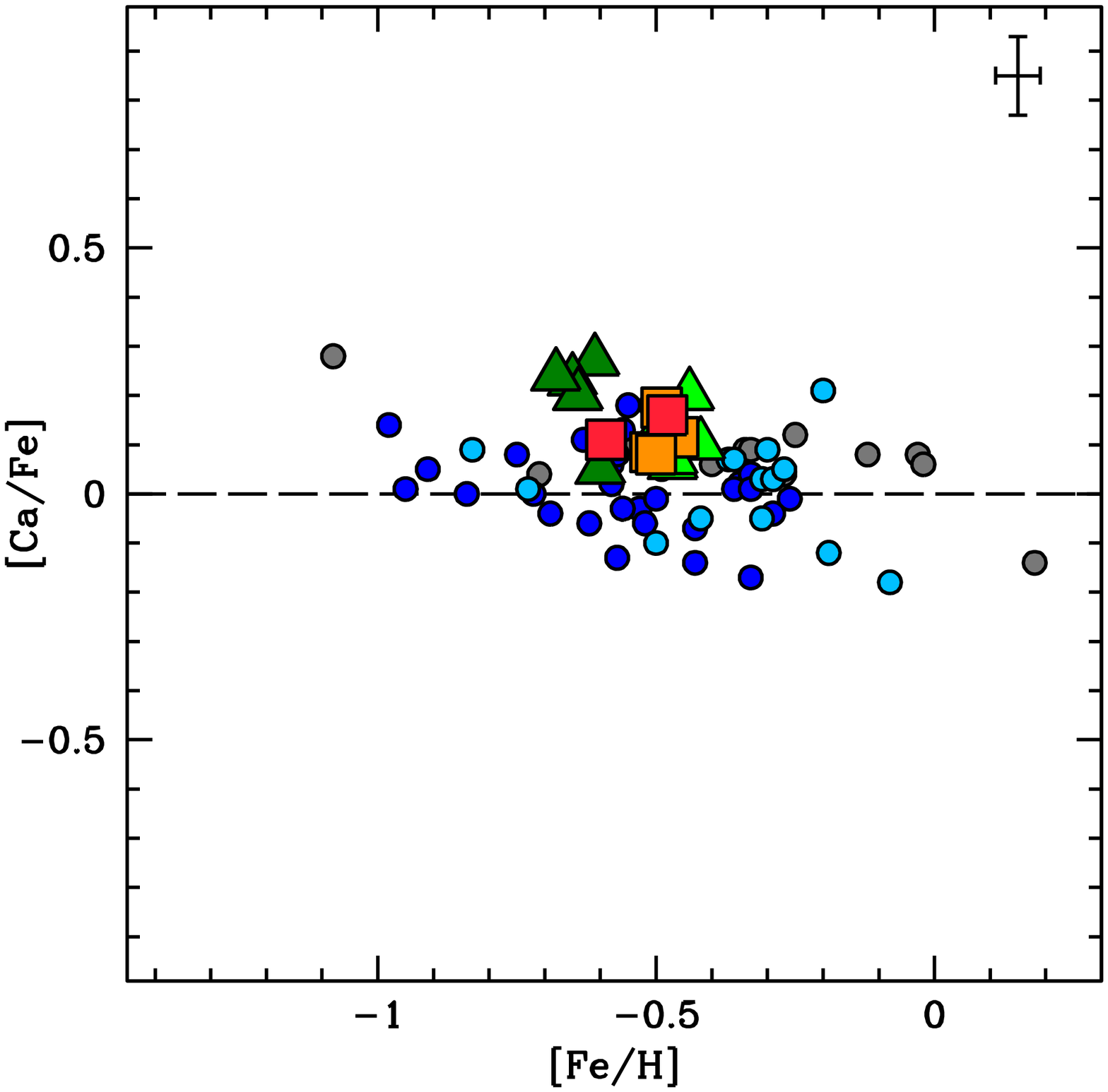}
\includegraphics[scale=0.29]{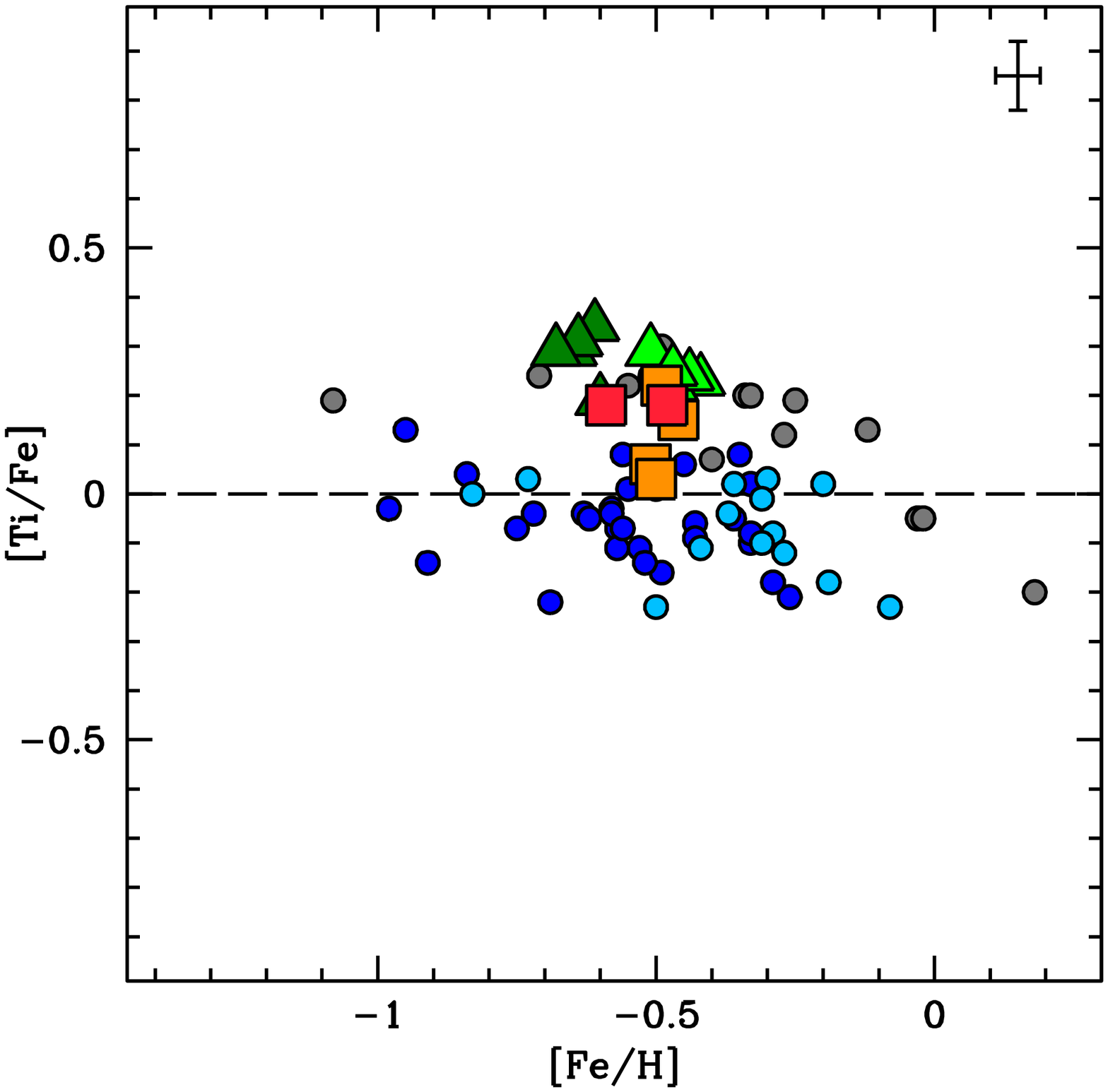}
\caption{Behavior of the $\alpha$ elements [Si/Fe], [Ca/Fe] and [Ti/Fe] abundance ratios (from panel left to panel right, respectively) as a function of [Fe/H] for NGC 5927 (light green triangles), NGC 6388 (orange squares), NGC 6441 (red squares), NGC 6496 (dark green triangles), with LMC (blue dots), Sgr (light blue dots) and MW (grey dots) as reference.}
\label{alfa}
\end{figure}

\begin{figure}
\centering
\includegraphics[scale=0.29]{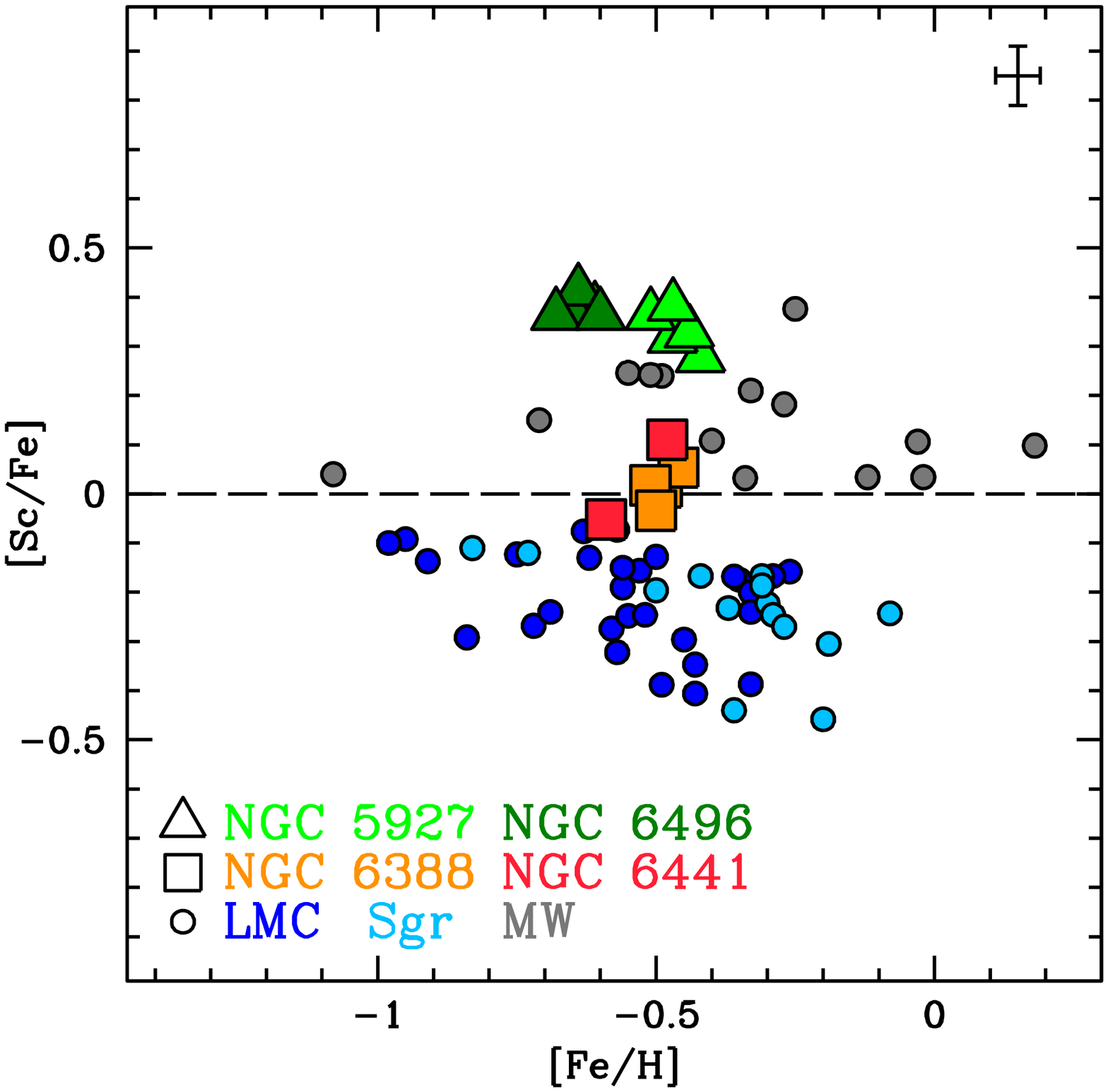}
\includegraphics[scale=0.29]{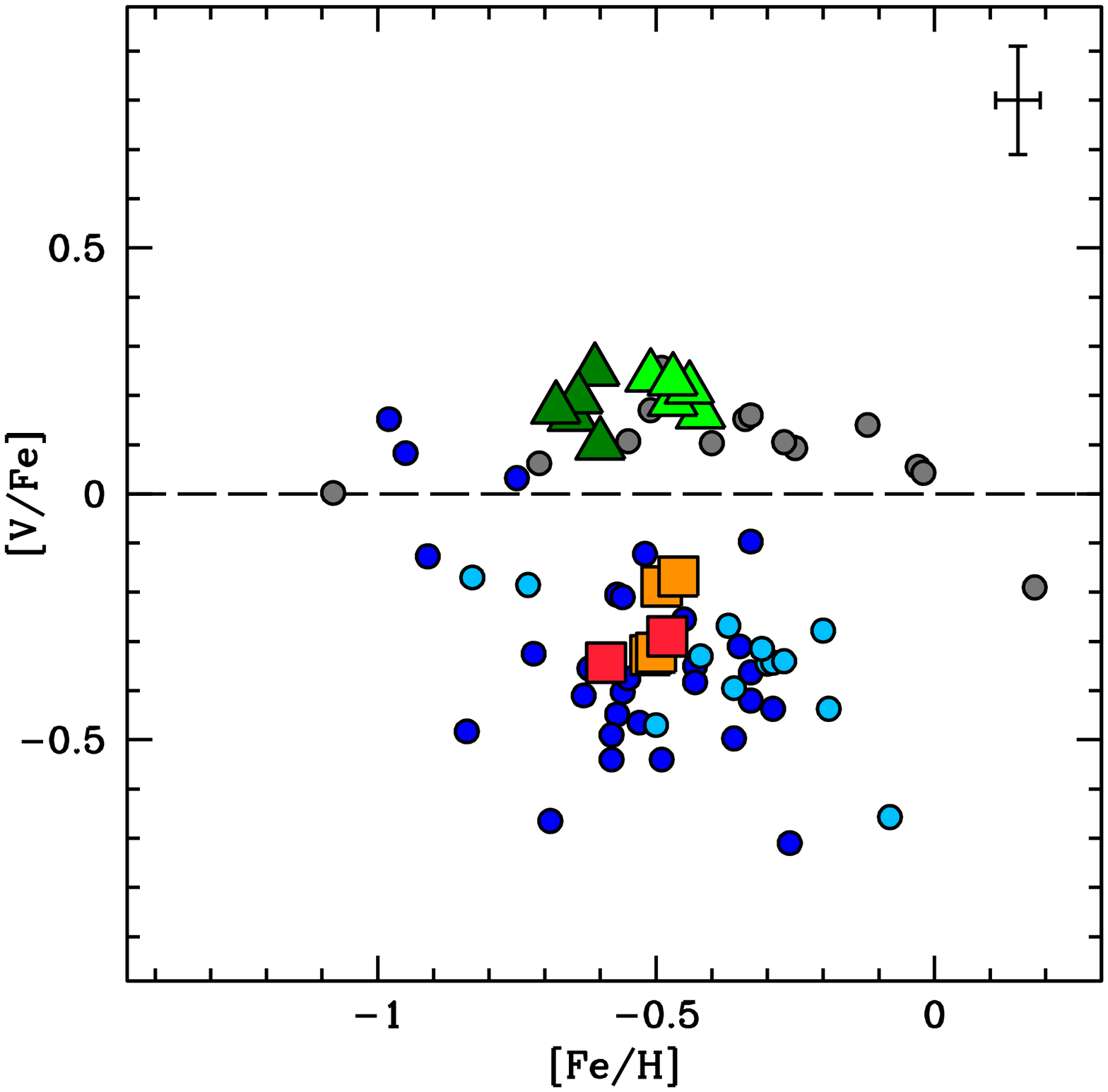}
\includegraphics[scale=0.29]{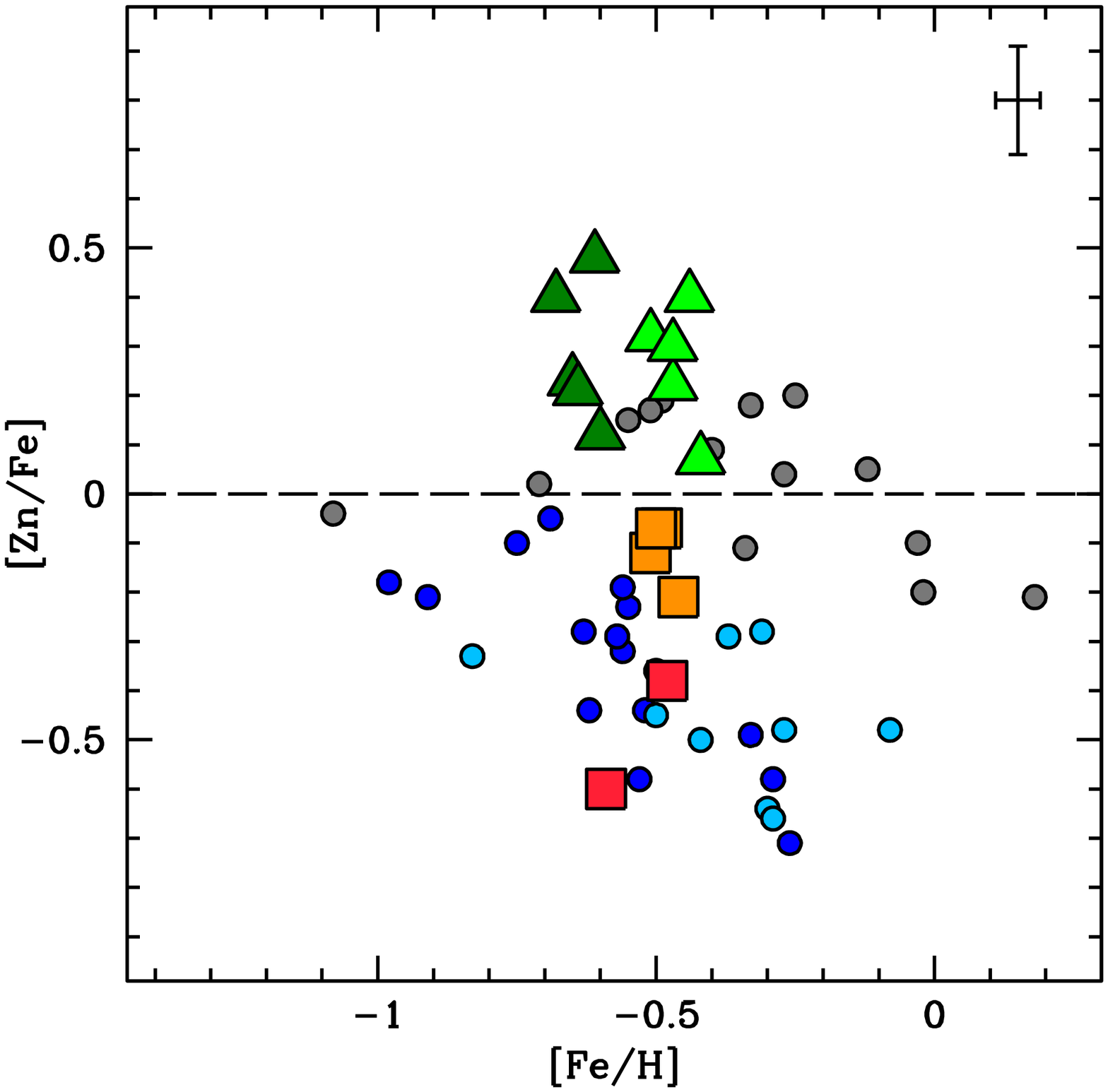}
\caption{Behavior of the iron-peak elements [Sc/Fe], [V/Fe] and [Zn/Fe] abundance ratios (from panel left to panel right, respectively) as a function of [Fe/H] for NGC 5927, NGC 6388,  NGC 6441, NGC 6496, LMC, Sgr and MW (same symbols of Fig. \ref{alfa}).}
\label{iron}
\end{figure}

\begin{figure}
\centering
\includegraphics[scale=0.29]{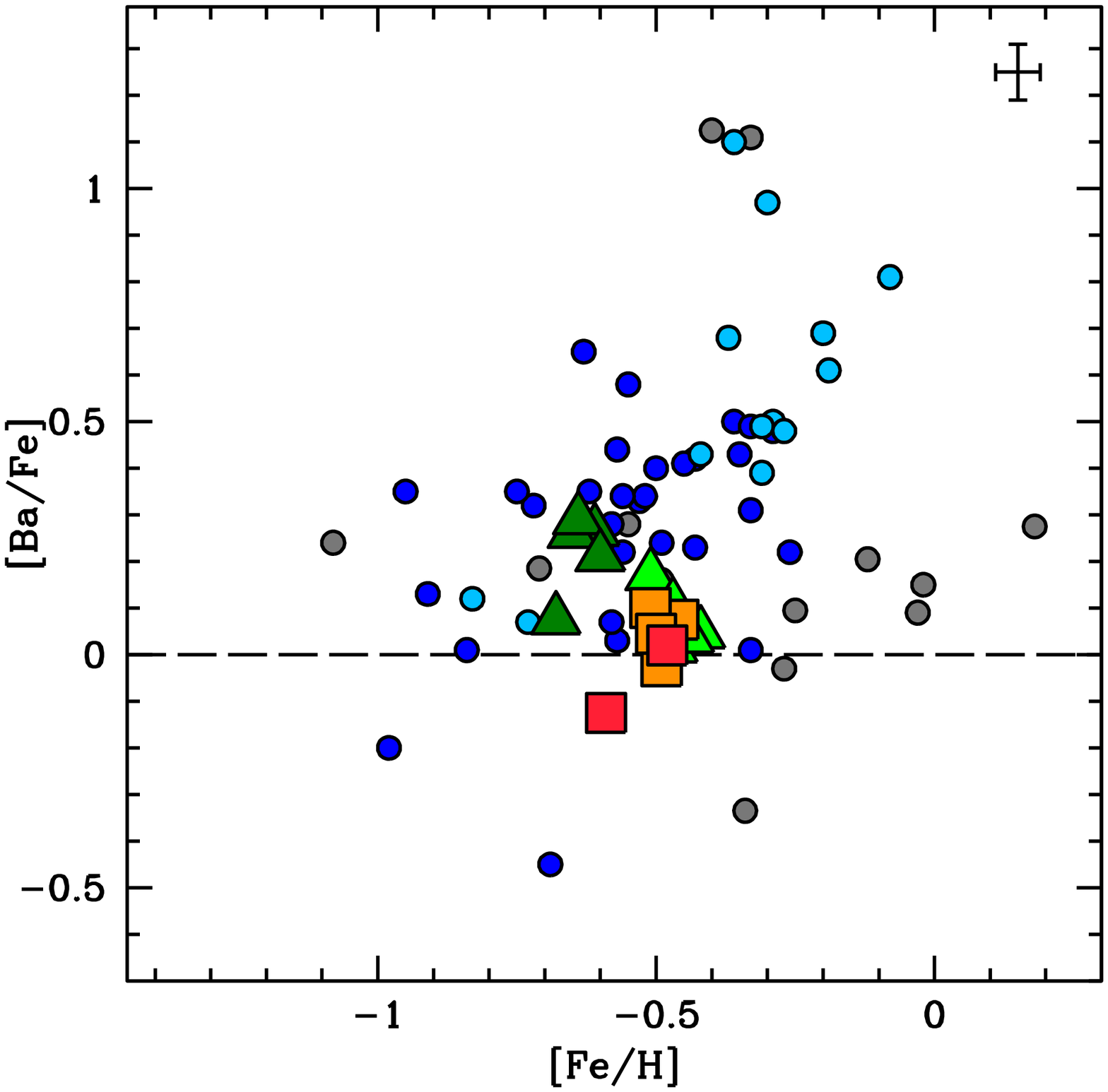}
\includegraphics[scale=0.29]{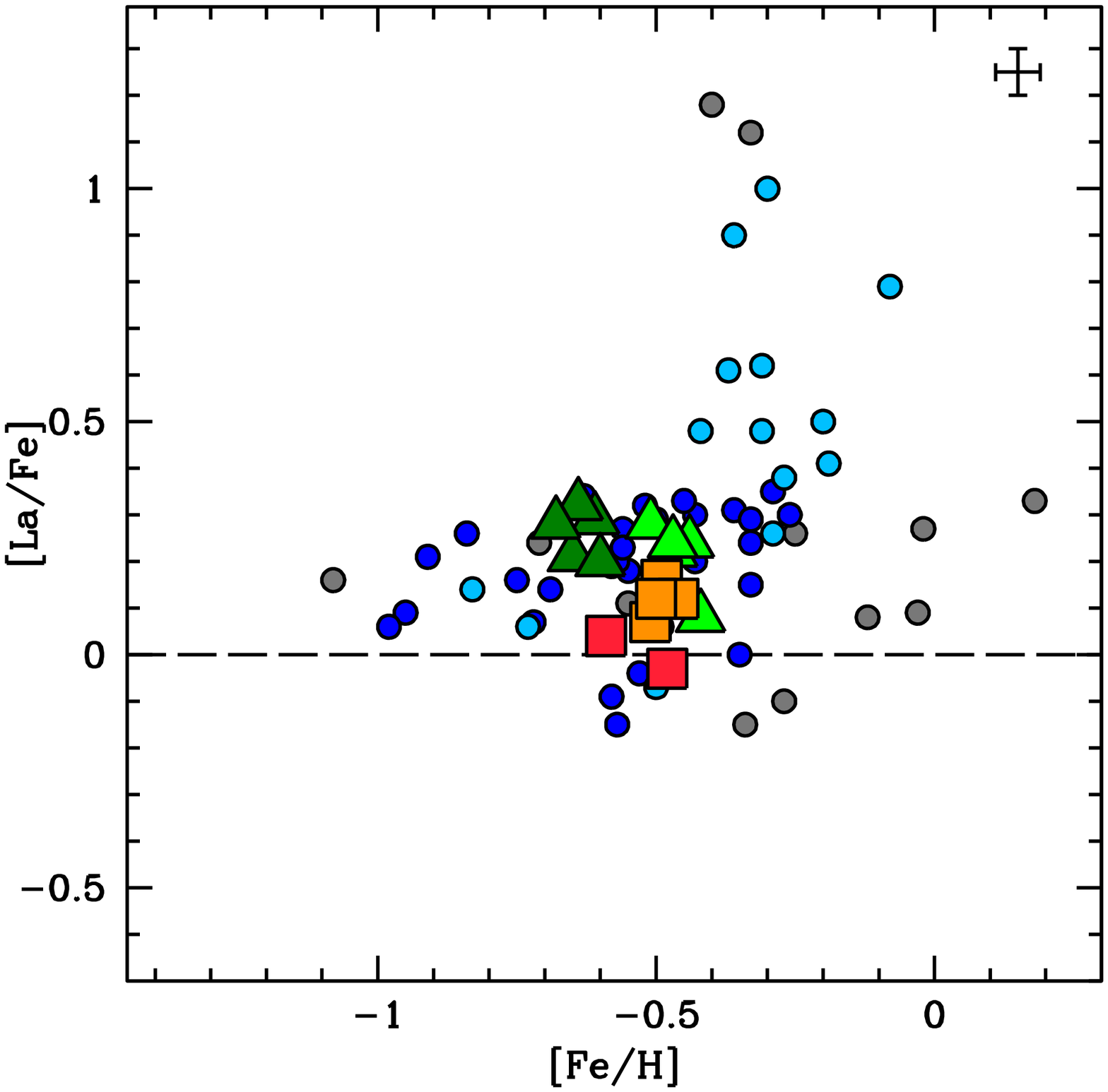}
\includegraphics[scale=0.29]{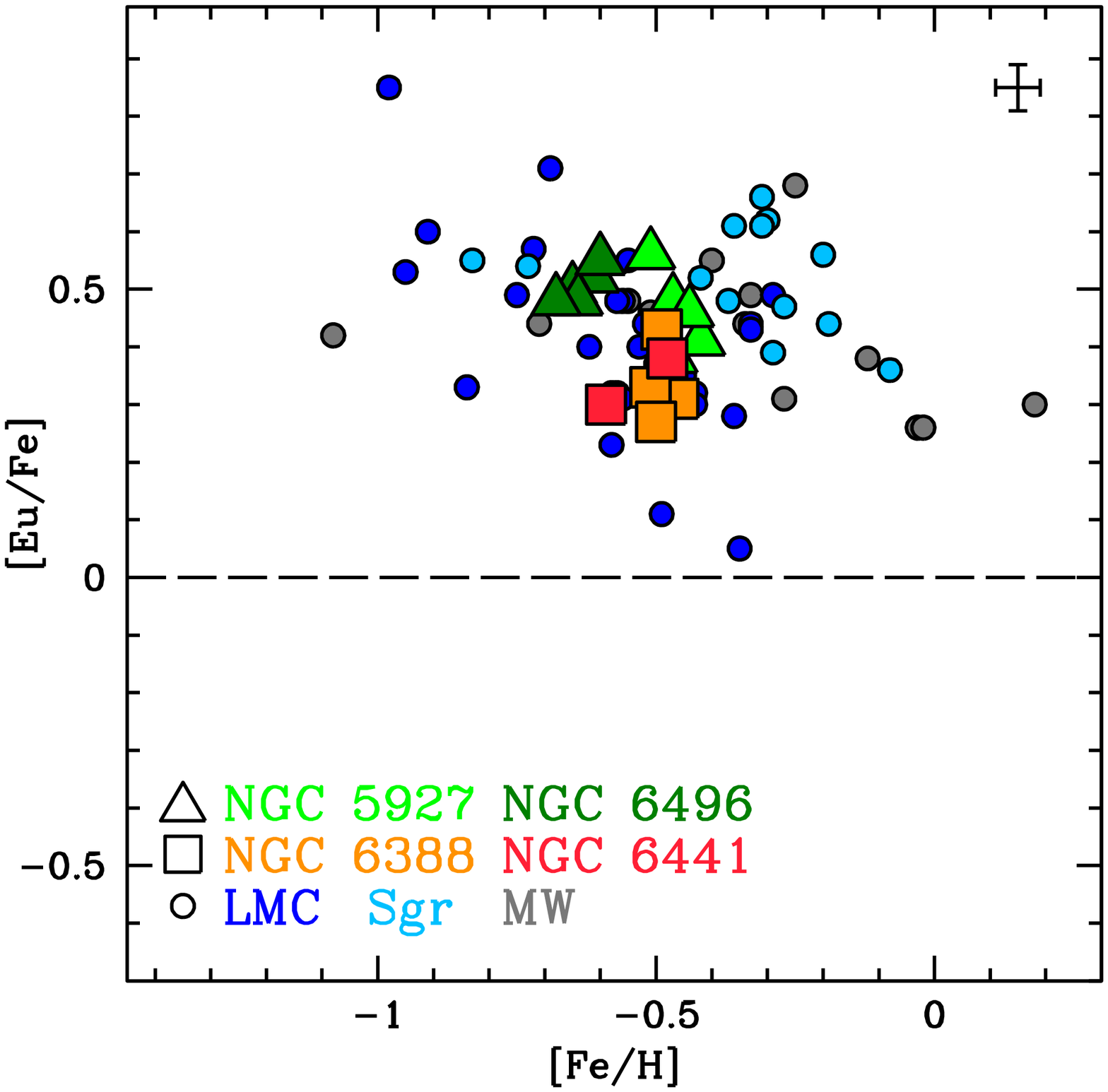}
\caption{Behavior of the neutron capture elements [Ba/Fe], [La/Fe] and [Eu/Fe] abundance ratios (from panel left to panel right, respectively) as a function of [Fe/H] for NGC 5927, NGC 6388,  NGC 6441, NGC 6496, LMC, Sgr and MW (same symbols of Fig. \ref{alfa}).}
\label{neutron}
\end{figure}

Figs.\ref{alfa}-\ref{neutron} show the measured abundance ratios for $\alpha$-, iron-peak and neutron-capture elements, 
as a function of [Fe/H] for the stars analysed in the four target clusters.
We can immediately appreciate that the $\alpha$-elements Si and Ca show similar abundance ratios in all
the four GCs. Slow (La and Ba) and rapid (Eu) neutron capture elements, in addition to Ti, show a marginal discrepancy, with NGC 6388 and NGC 6441 being under-abundant compared to NGC 5927 and NGC 6496 at 1-2 sigma level from the comparison between the mean abundance values and their standard deviation. On the other hand, a stark difference (at a significance level always larger than 3 sigma, up to $\sim$ 10 sigma) is found when considering the abundances of Sc, V and Zn. 
 In particular NGC 6388 and NGC 6441 have abundance ratios for these iron-peak elements significantly lower than those 
 measured in NGC 5927 and NGC 6496. 
 We stress that these differences cannot be attributed to some systematics in the chemical analysis
 because the assumptions in the analysis of all the GCs are the same 
 (i.e. the reference solar abundances, the atomic data, the model atmospheres, the method to derive 
 the atmospheric parameters), and we analyse stars of similar spectral type.
 Therefore, the origin of the different [Sc/Fe], [V/Fe] and [Zn/Fe] chemical abundance ratios must be intrinsic, due to a real difference in the chemical enrichment path followed by the gas from which the two pairs of clusters formed.

Interesting enough, the differences in these abundance ratios for the two pairs of GC, match well those measured by \citet{Minelli2021} between LMC/Sgr and MW field stars of similar metallicity (overplotted in Figs. \ref{alfa}-\ref{neutron} as small filled circles). In particular, NGC~6388 and NGC~6441 exhibit [Sc/Fe], [V/Fe] and [Zn/Fe] abundances similar to those measured in LMC/Sgr stars, while NGC~5927 and NGC~6496 have abundances similar to those of MW stars. We remark that according to many results in the literature \citep{Bensby2003, Bensby2017, Battistini2015, Duong2019, Griffith2021, Lucey2021} the iron-peak elements abundance ratios of MW disk and bulge stars are consistent with each other. All these works found abundance values similar to those of NGC 5927 and NGC 6496, but different from those of NGC 6388 and NGC 6441.
\cite{Minelli2021} interpret the low abundance ratios in LMC/Sgr stars in terms of a lower contribution from massive stars to the chemical enrichment, 
compared to that experienced by the MW. The reason for this would be that these elements are mainly produced by hypernovae, Type II supernovae or electron-capture supernovae with high-mass stellar progenitors. 
In particular, hypernovae (associated to stars more massive than $\sim$25-30 $M_{\odot}$) would produce most of Zn, without a sizeable contribution from Type Ia Supernovae \citep{Romano2010, Kobayashi2020}. 
Hence, the ratio [Zn/Fe] is expected to decrease significantly in galaxies with a low star formation rate, where the contribution by massive stars is reduced \citep{Yan2017,Jerabkova2018}.

In light of this finding, it is natural to conclude that both NGC 6388 and NGC 6441 should have formed from a gas poorly enriched by massive stars, at odds with what observed for the other two investigated clusters. 
Thus, the analysis presented here offers an independent confirmation that NGC 5927 and NGC 6496 formed in-situ, as already suggested by the kinematics \citep{Massari2019}, and identifies NGC 6388 and NGC 6441 as likely formed in an external environment, characterized by chemical enrichment histories influenced by a low star formation rate, and only later accreted by the MW.
It is interesting to note, that of the two clusters identified here as accreted, the kinematics analysis by \cite{Massari2019} indicated only NGC 6441 as an accreted cluster associated to the Kraken merger event \citep[see][]{Kruijssen2020}, while an unclear origin was indicated for NGC 6388. Therefore, in light of the similar chemical abundances of the two GCs, NGC 6388 should have formed from the same progenitor of NGC 6441 or at least from a system with a chemical enrichment history similar to that of Kraken.

In summary, the use of the iron-peak elemental abundances proposed by \cite{Minelli2021} has allowed us to shed light on the origin of the metal-rich GCs NGC 6338 and NGC6441,
indicating also NGC 6388 as a possible accreted cluster from a progenitor similar to Kraken in spite of the fact that its dynamical properties were not sufficient to unambiguously determine its birth place.

Unlike Zn, whose nucleosynthesis in stars is pretty well understood (see previous paragraphs), the detailed nucleosynthetic paths leading to the stellar production of Sc and V still deserve investigation \citep[see][for recent reappraisals from the observational and theoretical point of view, respectively]{Cowan2020, Kobayashi2020}. Notably, different initial conditions of exploding white dwarfs leading to type Ia supernovae may result in very different V yields \citep[e.g.][]{Shen2018, Leung2020} with sizable consequences on the predictions of chemical evolution models \citep{Palla2021} that have still to be fully explored. Our results clearly highlight the importance of Sc, V and Zn as chemical taggers and will hopefully inspire further theoretical work.

\acknowledgments
We thank the referee, Chris Sneden, for his useful comments and suggestions.
This research is funded  by the project "Light-on-Dark" , granted by the Italian MIUR
through contract PRIN-2017K7REXT.

\end{document}